\documentclass[a4paper,11pt]{article}
\usepackage{jheppub}
\notoc

\title{The use of a high intensity neutrino beam from the ESS proton linac for measurement of neutrino CP violation and mass hierarchy}

\author[a]{E~ Baussan,}
\author[a,*]{M.~Dracos,}
\author[b,*]{T.~Ekel\"of,}
\author[c]{E.~Fernandez~Martinez,}
\author[b]{H.~\"Ohman} 
\author[a]{and N.~Vassilopoulos}

\affiliation[a]{IPHC, Universit\'e de Strasbourg, CNRS/IN2P3, F-67037 Strasbourg, France}
\affiliation[b]{Department of  Physics and Astronomy, Uppsala University, Box 516, SE-75120 Uppsala, Sweden}
\affiliation[c]{Theory Divison, CERN, CH-1211 Geneva 23, Switzerland and Dpto. de F\'isica T\'eorica and Instituto de F\'isica T\'eorica UAM/CSIC, Universidad Aut\'onoma de Madrid, Cantoblanco E-28049 Madrid, Spain.}
\affiliation[*]{Corresponding Authors: marcos.dracos@in2p3.fr and tord.ekelof@physics.uu.se}

\abstract{
It is proposed to complement the ESS proton linac with equipment that would enable the production, concurrently with the production of the planned ESS beam used for neutron production, of a 5~MW beam of~10$^{23}$ 2.5~GeV protons per year in microsecond short pulses to produce a neutrino Super Beam, and to install a megaton underground water Cherenkov detector in a mine to detect $\nu_e$ appearance in the produced $\nu_\mu$ beam.
Results are presented of preliminary calculations of the sensitivity to neutrino CP violation and the mass hierarchy as a function of the neutrino baseline.
The results indicate that, with 8 years of data taking with an antineutrino beam and 2 years with a neutrino beam and a baseline distance of around 400 km, CP violation could be discovered at 5~$\sigma$ (3~$\sigma$) confidence level in 48\% (73\%) of the total CP violation angular range.
With the same baseline, the neutrino mass hierarchy could be determined at 3~$\sigma$ level over most of the total CP violation angular range.
There are several underground mines with a depth of more than 1000~m, which could be used for the creation of the underground site for the neutrino detector and which are situated within or near the optimal baseline range.
}
\keywords{neutrino, super beam, ESS, EUROnu, SPL}
\begin{document}
\maketitle

\section{Introduction}
The magnitude of CP violation as observed in the quark sector is not enough for the Standard Model to explain the dominance of matter over antimatter in the Universe. Following the discovery of neutrino oscillations, and thereby of neutrino mixing, it has become an important task to discover and measure CP violation in the neutrino sector, as this could be related to the generation of the matter dominance. 

Most of the parameters in the neutrino mixing matrix have by now been measured with good precision. The latest mixing angle discovered to be non-zero by several experiments \cite{0,1,2,3} was the mixing angle $\theta_{13}$. The currently most precise single measurement of sin$^22\theta_{13}$ is 0.089$\pm$0.010(stat)$\pm$0.005(syst) \cite{3}. The only two remaining undetermined neutrino oscillation parameters are the neutrino mass ordering and the CP violating phase angle~$\delta$. In the expression for the $\nu_\mu$$\to$$\nu_e$ probability,  sin$^22\theta_{13}$ multiplies the CP violating parameter. The relatively large measured value for  sin$^22\theta_{13}$ now opens the possibility to discover and measure CP violation in the neutrino sector using a high intensity ``conventional'' neutrino Super Beam experiment. As to the Neutrino mass hierarchy ordering, presently approved projects could only give an indication about this parameter. 

\begin{figure}[hbt]
\begin{center}
\setlength\fboxsep{0.6pt}
\setlength\fboxrule{0.6pt}
  \includegraphics[width=1\textwidth]{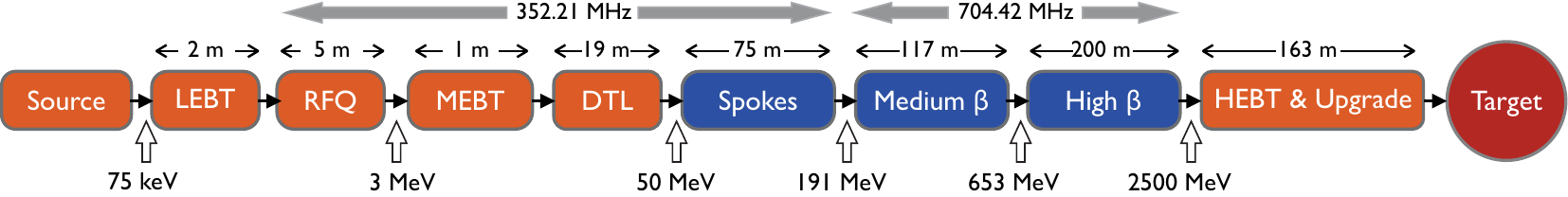}
\end{center}
\caption{  Schematic drawing of the 2.5~GeV proton linear accelerator of the European Spallation Source (ESS).\label{fig:1} }
\end{figure}

We propose to use the 2.5~GeV proton linear accelerator of the European Spallation Source (ESS) \cite{4} in Lund, Sweden (Fig. \ref{fig:1}), currently under construction, to generate a low energy high intensity $\nu_\mu$ beam, similar to the CERN Super Beam from the CERN Super Proton Linac \cite{5} proposed by the FP7 EUROnu Design Study \cite{6}. The power of the ESS proton beam for neutron production will be 5~MW. The ESS proton linac will thereby, when it is taken into operation in 2019, become the most powerful neutron source in the world and remain so for a number of years.

The present proposal is to increase the power generation capacity of the ESS linear accelerator by another 5~MW to make possible the production of~10$^{23}$ protons per year for neutrino production, concurrently with the production of protons for neutron spallation. 

As detector for the appearing $\nu_e$ we propose to use a large water tank Cherenkov detector located underground in a mine at a depth of at least 1000~m (3000~m water equivalent). In this paper we estimate what range of CP phase angles would allow for a discovery of CP violation and address, in particular, the question of at which distance from the neutrino source the detector should be placed to maximize this range, and also what sensitivity to the mass hierarchy could be achieved.

\section{The use of the ESS proton linac to generate a neutrino beam}
The ESS proton linac is about 500~m long and uses normal conducting drift tubes and superconducting spoke cavities at 352~MHz and superconducting elliptical cavities at 704~MHz to accelerate 14 equally spaced 3~ms long proton pulses per second. The long time of 70~ms between pulses is motivated by that one wants to employ time of flight methods to measure the relatively slowly moving spallation neutrons in the experiments. The present proposal is to inject 3~ms long H$^-$ ion pulses in the 70~ms time gaps between the proton bunches and accelerate these concurrently with the protons. After acceleration the H$^-$ ion beam will be separated from the proton beam and injected into a proton storage ring where the H$^-$ ions are successively stripped of their electrons and accumulated in the ring. The ring is then emptied in one turn, thus making the delivery of a proton pulse of a length of few microseconds possible. 

The thus compressed proton pulses are aimed at a neutrino target surrounded by a neutrino horn magnet. Depending on the polarity of the horn magnet, the produced positively or negatively charged pions will be focused in the forward direction. Downstream of the target there is a decay tunnel of length of order 20 m. This is long enough for most of the pions to decay into muon neutrinos and muons and at the same time sufficiently short to prevent that a significant amount of electron neutrinos are created from muon decays. At 2.5~GeV proton energy kaon production is considerably reduced compared to higher energies and does not produce significant electron neutrino background. There will be no background from tau neutrinos as the energy will be below the threshold for tau production (a significant fraction of mu neutrinos oscillates to tau neutrinos). The parameters of the target station installation used for the present study are the same as those used in the study made by EUROnu of the Super Proton Linac (SPL) Super Beam. 

What needs to be added to the already planned ESS proton linac equipment is thus a H$^-$ source, an additional 5~MW radiofrequency power source, an accumulator ring, a neutrino target with horn and a decay tunnel. A megaton underground water Cherenkov detector is needed for the detection and identification of the neutrinos at large distance from the neutrino source. A smaller neutrino detector located near the neutrino source is required for the determination of the neutrino flux. The investment cost for an additional 5~MW radiofrequency power source is estimated to be of the order of 100~MEUR which may be compared to the investment cost for the whole ESS linac, which is of the order of 700~MEUR. All of these items will require extra space at the ESS site in Lund. For the present proposal to remain a viable option in future, it is important that the current planning of the installation of the ESS proton linac, spallation target and experimental areas on the ESS site be such that space for the above items can be made available. It may be noted that there has already been interest expressed from the spallation neutron community to have an accumulator ring added to the current ESS project to be able to produce short neutron pulses.

\section{The neutrino detector and the simulation code}
\begin{figure}[hbt]
\begin{center}
\setlength\fboxsep{0.6pt}
\setlength\fboxrule{0.6pt}
\fbox{
  \includegraphics[width=0.5\textwidth]{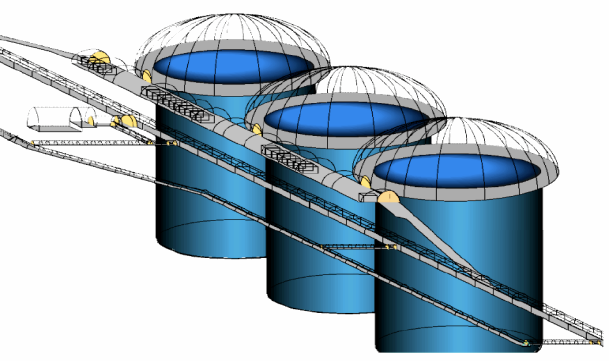}}
\end{center}
\caption{The MEMPHYS water Cherenkov detector of a total fiducial mass 440~kton consisting of three cylindrical modules of 65~m high and 65~m diameter and with 3x81000 12'' photomultipliers mounted on the walls, providing 30\% geometrical coverage. \label{fig:2}}
\end{figure}

Most of the neutrinos derived from the 2.5~GeV proton beam will have energies in the range 200-500~MeV. For such comparatively low energies the rate of inelastic events is limited, implying that it is sufficient to measure only the outgoing charged lepton in the event. Furthermore, the neutrino cross-section is lower than at higher energies, leading to the requirement of a comparatively large target volume. For the present proposal has been adopted a water Cherenkov detector with a fiducial volume of 440~kt of the same type as the MEMPHYS detector \cite{7} (see Fig. \ref{fig:2}) planned for the Fr\'ejus site with the CERN SPL as neutrino source, rather than a liquid scintillator or a liquid Argon detector, which are more expensive by unit volume but are needed for the reconstruction of inelastic events at higher energies, where such events are more frequent.

For the calculations presented below we have used the simulation code GLOBES \cite{8} elaborated for the MEMPHYS detector with the CERN SPL as neutrino source. For the neutrino flux calculation, the proton energy has been set to 2.5~GeV and the proton flux to~10$^{23}$ per year. In the present study the neutrino target parameters (horn current, target position in the horn and the length and diameter of the decay tunnel), optimized for the 4.5~GeV proton energy of SPL, have so far not been reoptimized for the ESS 2.5~GeV energy. The smaller aperture neutrino detector near the neutrino source, that will be needed to measure the flux of neutrinos, has not been investigated in the present study.

\section{Optimization of the baseline length}
\begin{figure}[hbt]
\begin{center}
  \includegraphics[width=0.6\textwidth]{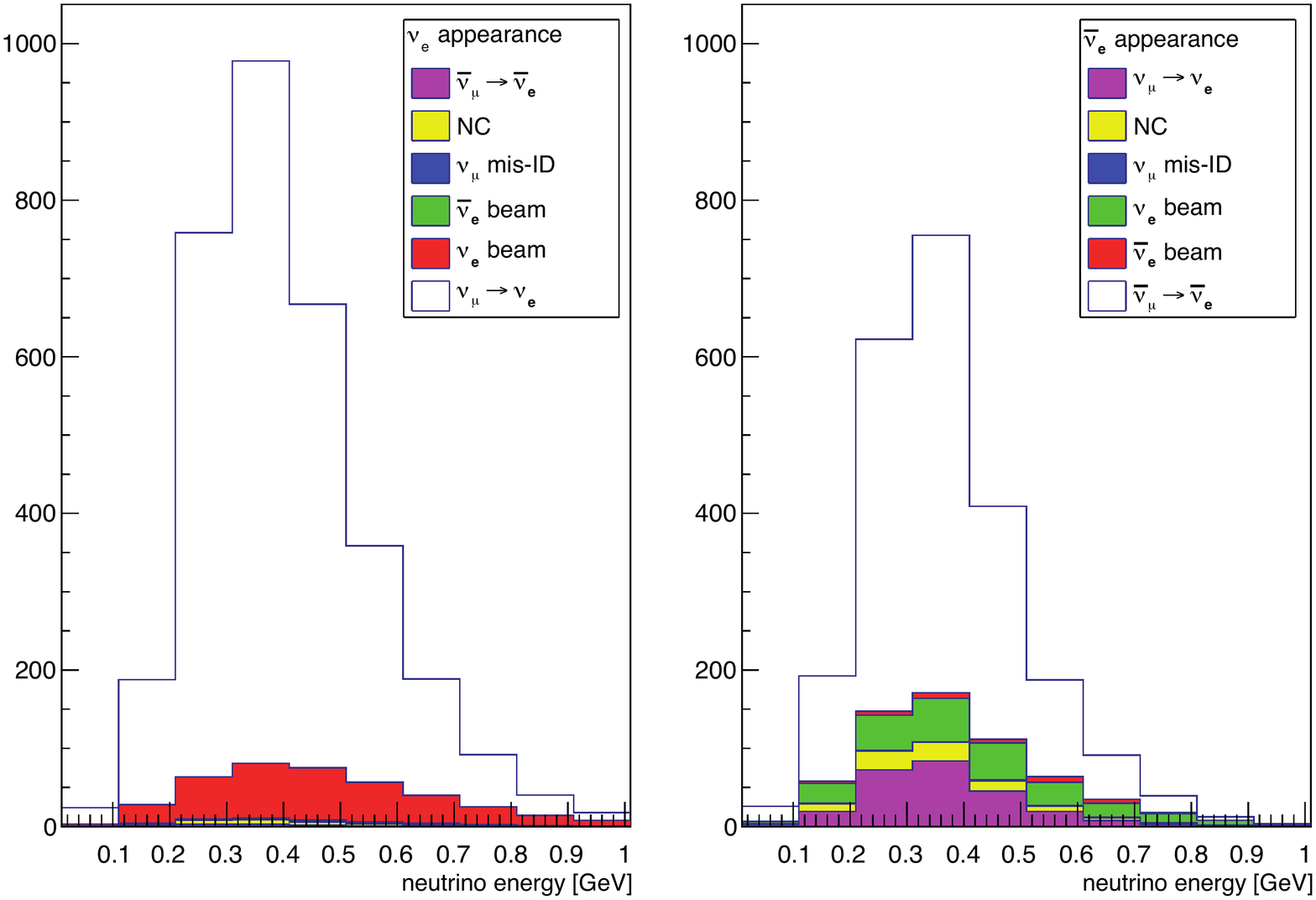}
\end{center}
\caption{Electron neutrino and antineutrino energy spectra including background contribution as detected by MEMPHYS for 2 years of neutrino running (left) plus 8 years of antineutrino running (right) and a baseline length of 150~km. \label{fig:3}}
\end{figure}

In order to evaluate numerically the discovery potential of CP violation for different baseline lengths $L$ and for different values of the CP phase~$\delta$, we have used the GLOBES code to simulate the neutrino oscillation and the detection of the neutrinos in the MEMPHYS detector, varying $L$. The parameter values used in the GLOBES calculation are:~$\Delta $m$^2_{12}$ = 7$\times 10^{-5}$ eV$^2$,~$\Delta $m$^2_{31}$ = 2.43$\times 10^{-3}$ eV$^2$ (normal or inverted hierarchy), $\theta_{12}$ = 0.59 and $\theta_{23}$ = $\pi$/4.  For $\theta_{13}$  the Daya Bay result  sin$^22\theta_{13}$ = 0.089 has been used. The neutrino mass hierarchy is not assumed to be known. These parameters are included in the fit assuming a prior knowledge with an accuracy of 10\% for $\theta_{12}$ and $\theta_{23}$, 5\% for~$\Delta $m$^2_{31}$ and 3\% for~$\Delta $m$^2_{12}$ at 1~$\sigma$ level. The error in  sin$^22\theta_{13}$ has been set to 0.011 (= the Day Bay statistical and systematic errors added in quadrature). A systematic uncertainty of 5\% was assumed for the neutrino flux normalization and a 10\% systematic uncertainty for the background. The data collection period used is 2 years of neutrino running plus 8 years of antineutrino running. Fig. \ref{fig:3} shows the electron and antielectron neutrino spectra as detected by MEMPHYS for a baseline length of 150 km, which is before the first maximum is attained. Also shown are the contributions from various background sources. The neutrino mean energy is approximately 350~MeV with a large FWHM of about 300~MeV and a tail towards higher energies.
 
\begin{figure}[hbt]
\begin{center}
  \includegraphics[width=0.6\textwidth]{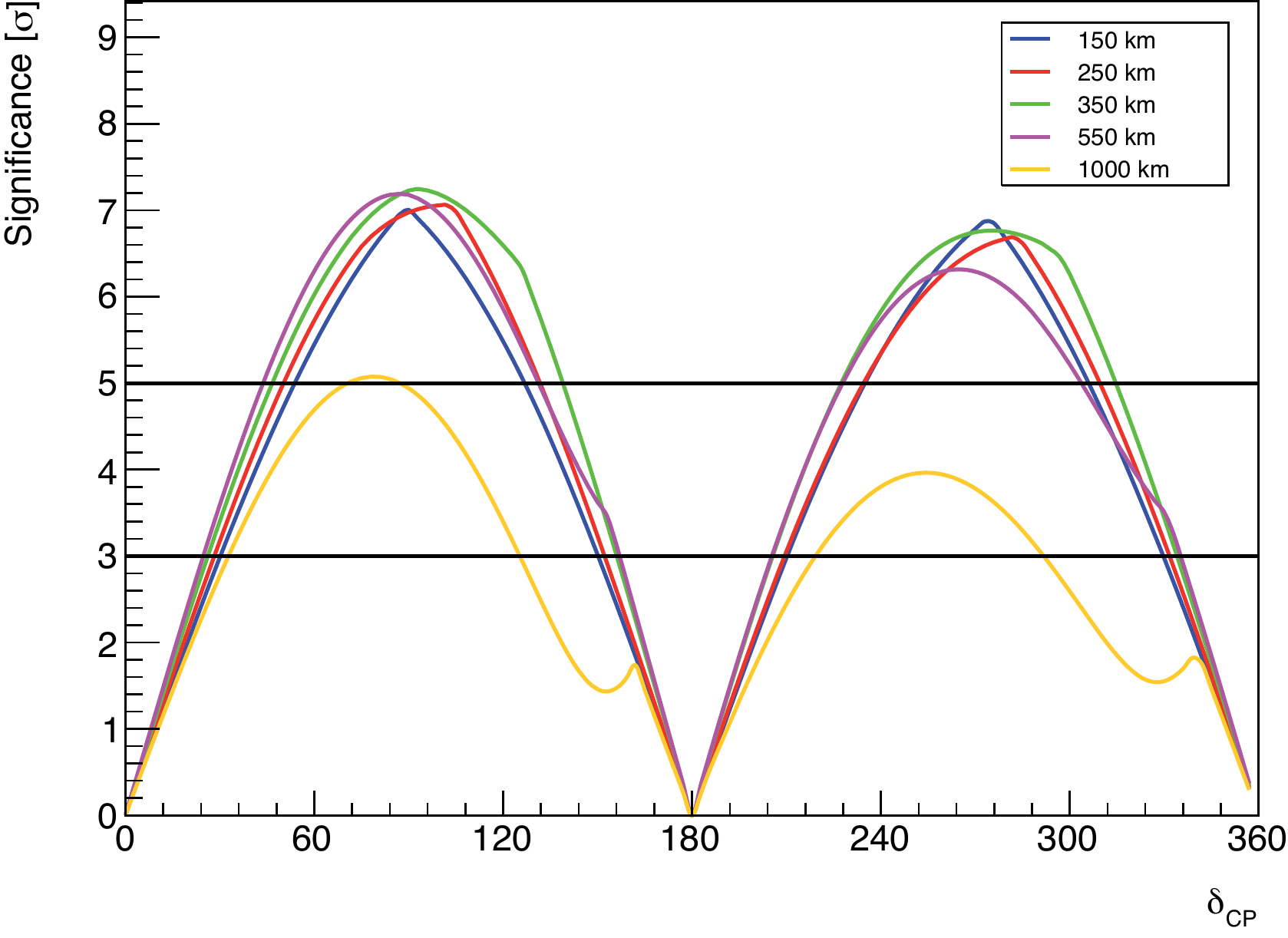}
\end{center}
\caption{The significance in terms of number of standard deviations $\sigma$ with which CP violation could be discovered for~$\delta$-values from 0$^{\circ}$ to 360$^{\circ}$ and for $L$ = 150, 250, 350, 550 and 1000 km. \label{fig:4}}
\end{figure}

\begin{figure}[hbt]
\begin{center}
  \includegraphics[width=0.6\textwidth]{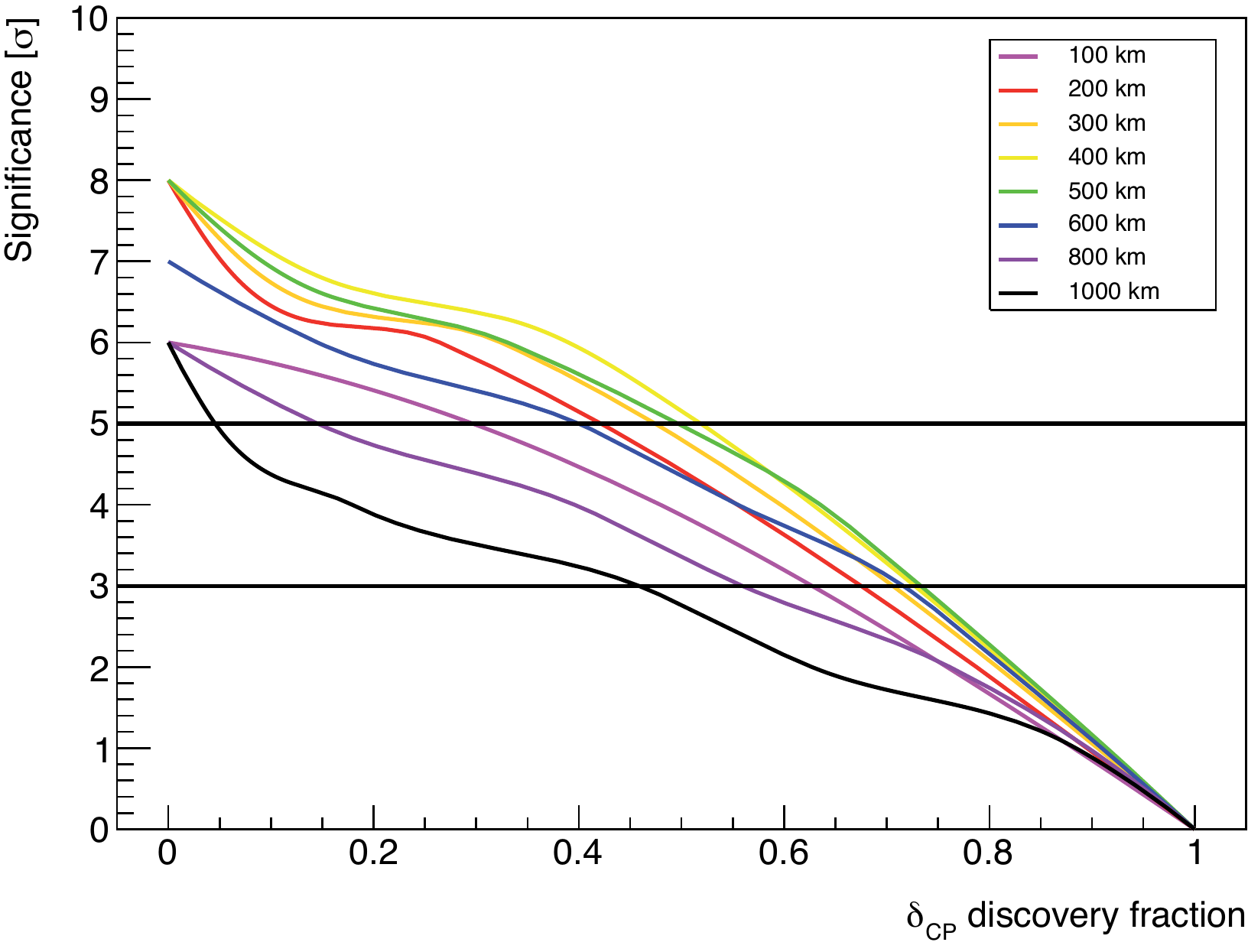}
\end{center}
\caption{ The significance in terms of number of standard deviations $\sigma$ with which CP violation can be discovered as function of the fraction of the full~$\delta$ range for $L$ = 100, 200, 300, 400, 500, 600 800 and 1000 km. \label{fig:5}}
\end{figure}

\begin{figure}[hbt]
\begin{center}
  \includegraphics[width=0.6\textwidth]{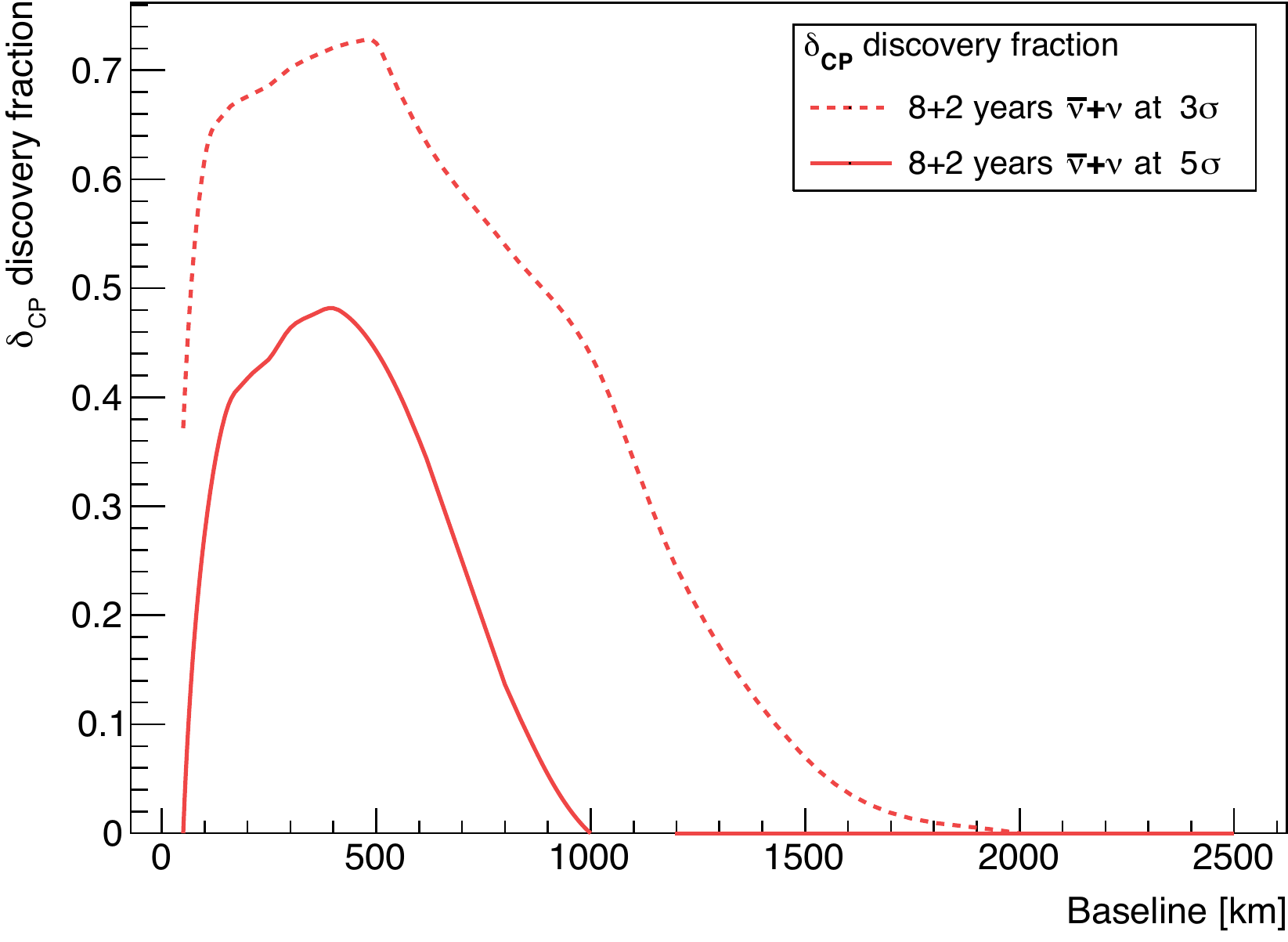}
\end{center}
\caption{ The fraction of the full~$\delta$ range as function of the baseline length in km. The lower (upper) curve is for CP violation discovery at 5~$\sigma$ (3~$\sigma$) significance \label{fig:6}.}
\end{figure}

The GLOBES code has been used to calculate the number of detected electron neutrinos and antineutrinos for values of~$\delta$ ranging from 0 to 360 degrees and for different baseline lengths $L$. For fixed baseline length $L$ and for each value of delta the number of events was generated and the $\chi^2$ was computed for delta = 0 and~$\delta$ = $\pi$ in the two neutrino mass hierarchy orderings, marginalized over all other oscillation parameters within their corresponding priors. The smallest of these four values (the best fit to a CP conserving value) was recorded. The square root of this number provided the significance in terms of standard deviations $\sigma$ with which CP violation could be discovered. This calculation was carried for a series of different baseline lengths. In Fig. \ref{fig:4},  \ref{fig:5} and  \ref{fig:6} different projections of the results of these calculations are displayed.

In Fig. \ref{fig:4}  is shown the significance in terms of number of standard deviations $\sigma$ with which CP violation could be discovered as a function of the value of~$\delta$ from 0$^{\circ}$ to 360$^{\circ}$. The different curves represent different distances $L$ = 150, 250, 350, 550 and 1000 km. The two horizontal lines have been drawn at the significance levels 3~$\sigma$ and 5~$\sigma$. 

In Fig. \ref{fig:5} is shown the significance in terms of number of standard deviations $\sigma$ with which CP violation can be discovered as function of the fraction of the full~$\delta$ range 0$^{\circ}$-360$^{\circ}$ within which CP violation can be discovered. The different curves represent different distances $L$ = 100, 200, 300, 400, 500, 600, 800 and 1000 km. The two horizontal lines have been drawn at the significance levels 3~$\sigma$ and 5~$\sigma$.

In Fig. \ref{fig:6} is shown the fraction of the full~$\delta$ range 0$^{\circ}$-360$^{\circ}$ within which CP violation can be discovered as function of the baseline length in km. 
According to the results of these calculations the fraction of the full~$\delta$ range 0$^{\circ}$-360$^{\circ}$ within which CP violation can be discovered at 5~$\sigma$ (3~$\sigma$) significance is above 42\% (68\%) in the range of baseline distances from 200 to 550~km and has the maximum value of 48\% (73\%) at around 400~km (500~km) baseline distance.

\begin{figure}[hbt]
\begin{center}
  \includegraphics[width=0.58\textwidth]{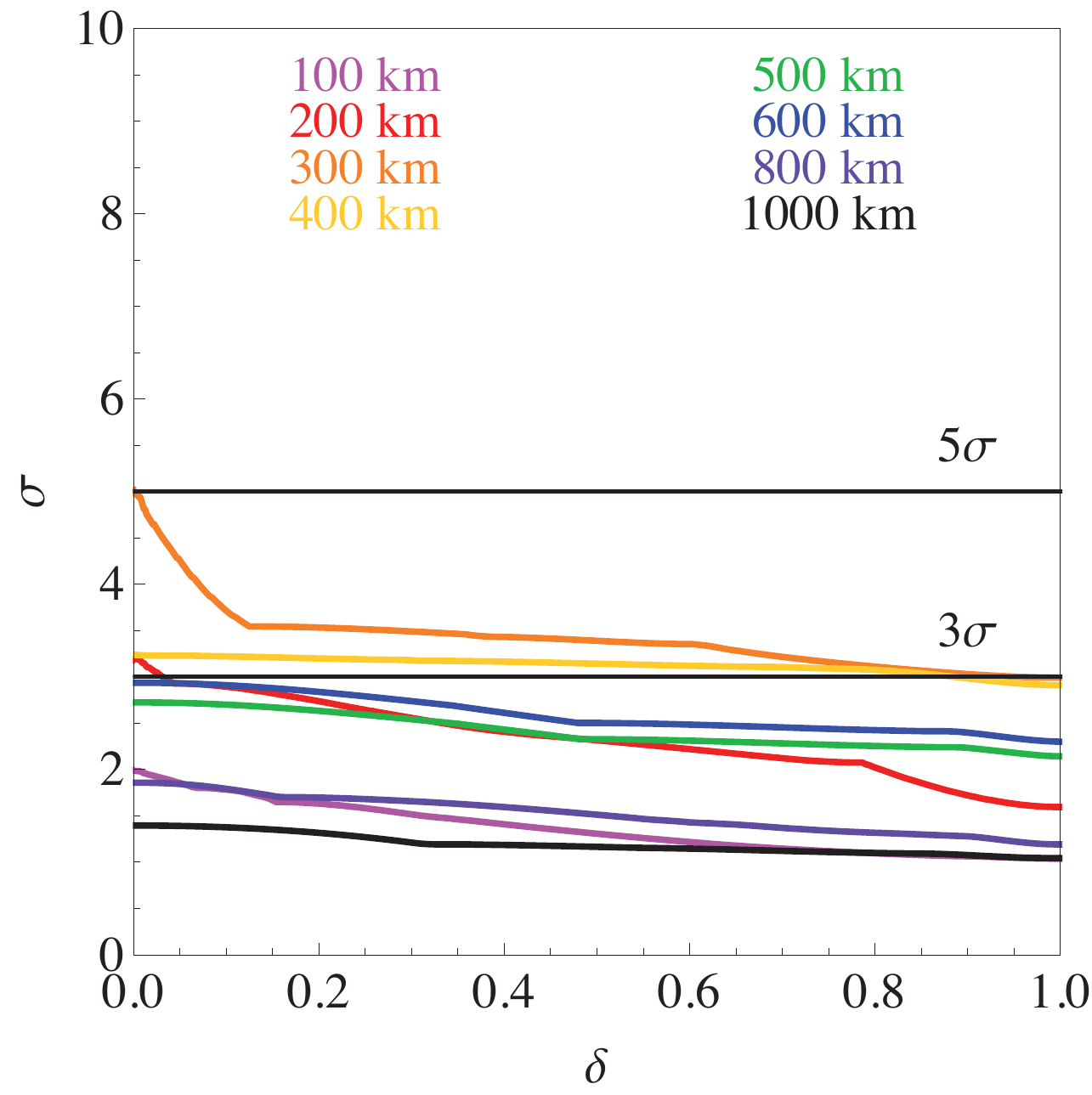}
\end{center}
\caption{ The significance in terms of $\sigma$ of the determination of the neutrino mass hierarchy as function of the fraction of the full~$\delta$ range 0$^{\circ}$-360$^{\circ}$. \label{fig:7}}
\end{figure}

To investigate the significance with which the neutrino mass hierarchy can be determined the analysis was carried out for a series of~$\delta$ values between 0$^{\circ}$ and 360$^{\circ}$ using both normal and inverted hierarchy and evaluating the significance of the difference between the two cases. The result is shown in Fig. \ref{fig:7}.
According to these results the mass hierarchy can be determined at 3~$\sigma$ significance within a large part of the full~$\delta$ range for baseline distances 300~km and 400~km but not for any baseline distance at 5~$\sigma$ significance. However, if combining these Super Beam neutrino measurements with the atmospheric neutrino measurements that can be concurrently made with the same detector \cite{9}, it should be possible - owing to MEMPHYS' very large volume - to obtain results for the mass hierarchy of higher significance than shown in Fig. \ref{fig:7}.

As already mentioned, the parameters of the neutrino target and horn, of the decay tunnel and of the neutrino detector have been taken from the EUROnu study for the SPL/Fr\'ejus Super Beam project, for which the proton beam energy is 4.5~GeV and the baseline length is fixed at 130~km. The next step in the current analysis will be to optimize these parameters for the proton beam energy of 2.5~GeV and different baseline lengths with the aim to further improve the physics performance for CP violation discovery and mass hierarchy determination.

Finally it may be noted that the powerful proton driver and accumulator based on the ESS linac as discussed in the present proposal could possibly, at a later stage, be used for the realization of a Neutrino Factory, which would allow to substantially enhance the performance for neutrino measurements.

\section{Available deep underground mines located within or near the optimal baseline length range}
\begin{figure}[hbt]
\begin{center}
  \includegraphics[width=1\textwidth]{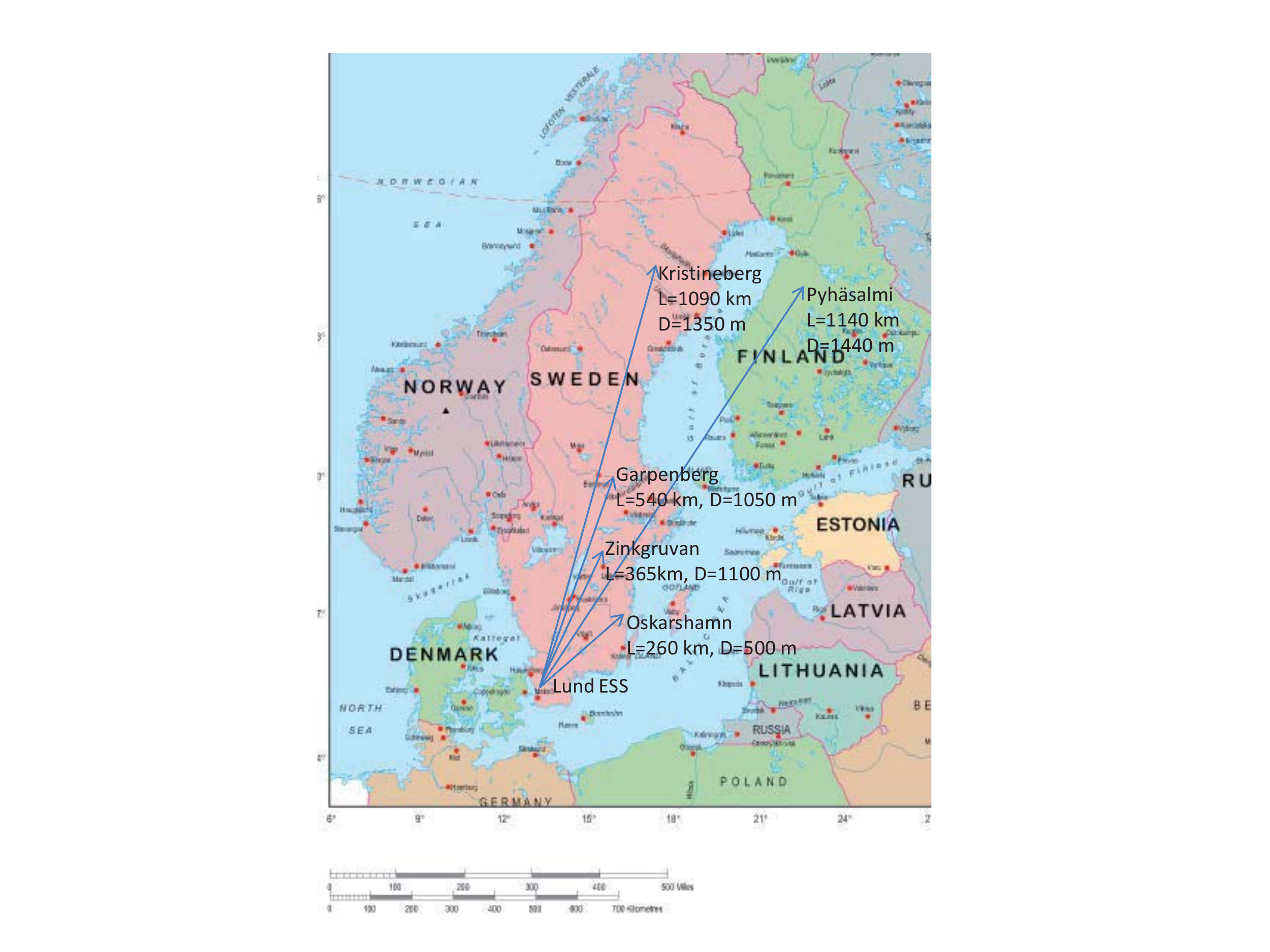}
\end{center}
\caption{ Location of the ESS site and the sites of several deep ($>$ 1000 m) underground mines. The distance ($L$) from ESS-Lund and the depth ($D$) of each mine is indicated below the name of the mine.\label{fig:8}}
\end{figure}

In order to protect the large neutrino detector from the cosmic ray background, the far detector needs to be located underground at a level of at least 1000~m (3000~m water equivalent). This is important in order to be able to carry out what would be the full research program of a MEMPHYS detector which comprises measurement of Super Beam neutrinos, proton decay, atmospheric neutrinos, supernovae neutrinos and geoneutrinos. The total volume of the detector used for the present calculations is of the order of 650000 m$^3$. The cost of the excavation and lining work of such a volume in the rock at 1000~m depth is around 100~MEUR. This cost assumes that there is already a large transport shaft down to this level and the infrastructure required for creating such an underground hall. Such infrastructure is normally available in mines. Fig. \ref{fig:8} shows the position in Sweden, the distance from the ESS site in Lund and the depth of several mines that are (or could be made) deeper than 1000~m.

The two active mines that are nearest to the ESS site is Zinkgruvan which is 1100~m deep and situated 365~km from Lund and Garpenberg, which is 1240~m deep and situated 540~km from Lund.
The closest inactive mine is St\"allberg, which is 1050~m deep and situated 490~km from Lund.
The exploitation of this mine ceased only about 35 years ago but its infrastructure could possibly be reactivated. 
Another, even nearer underground site is offered by the Oskarhamn nuclear waste depository site, which is situated at a distance of 260~km.
This depository is only 500~m deep, but can be extended down to below 1000~m.
There are more active and inactive mines further north in Sweden. Among the active mines there are Renstr\"om 1240~m deep and Kristineberg 1350~m deep, both at 1090~km from Lund.
The Pyhs\"almi mine in Finland, which has been studied as detector site for a long baseline beam from the SPS accelerator at CERN, is 1440~m deep and situated 1140~km from Lund.
According to the current status of our optimization calculations, Oskarshamn, Zinkgruvan and Garpenberg lie within the optimal baseline range for CP violation measurements. 

\section{Conclusion}
The currently planned and approved European Spallation Source proton linac will be ready in 2019.
Providing it with an extra H$^-$ source, an additional 5~MW radiofrequency power source, an accumulator ring, a neutrino target with horn and a decay tunnel, would make possible the production of a neutrino beam of about 350~MeV mean energy derived from~10$^{23}$, 2.5~GeV protons on target per year in concurrent operation with spallation neutron production.
The investment cost for upgrading the ESS linac to produce an extra 5~MW beam is modest compared to the cost to build a new separate proton driver of the same power.

Such a neutrino beam has the potential to become, during the next decade, the most intense neutrino beam in Europe and maybe also in the world.
The present preliminary study is based on Monte Carlo simulation of the generation and $\nu_\mu$$\to$$\nu_e$ oscillation in a neutrino beam produced with the use of the ESS proton linac and of the detection of $\nu_e$ using a very large Water Cherenkov detector of the MEMPHYS type. 
In the study has been determined the range in which the CP violating phase~$\delta$ in the neutrino sector could be discovered.
The preferred range of distances from the neutrino source to the detector site, within which a comparatively high potential for CP violation discovery is found, is between 200 and 550~km.
The results indicate that with 8 years of data taking with an antineutrino beam and 2 years with a neutrino beam up to 48\% (73\%) of the total CP violation~$\delta$ angular range could be covered at 5~$\sigma$ (3~$\sigma$) level at the optimal baseline distance of around 400~km. 
With the same baseline distance, the neutrino mass hierarchy could be determined at 3~$\sigma$ level over most of the total~$\delta$ angular range without extra optimization and without taking into account atmospheric neutrinos that would improve significantly this performance.

There are several underground mines with a depth of more than 1000~m, which could be used to facilitate the creation of the underground site for the neutrino detector and which are situated within or near the optimal baseline range. Further optimization, e.g. of the target station parameters and also of the detector design, are planned to improve on the sensitivity of the measurements of neutrino CP violation and to progress further with the determination of the optimal baseline distance.

\end{document}